\documentstyle[epsf,eqsecnum,floats,preprint,aps,epsfig]{revtex}

\input epsf
\tighten
\def\bk{{\bf k}}
\def\bp{{\bf p}}
\def\bx{{\bf x}}

\def\cL{{\cal L}}
\def\cO{{\cal O}}

\def\cPR{{\cal P}_{\cal R}}

\def\mpl{M_{\rm Pl}}
\def\half{\frac{1}{2}}

\begin{document}
\newcommand{\bm}[1]{\mbox{\boldmath{$#1$}}}
\newcommand{\be}{\begin{equation}}
\newcommand{\ee}{\end{equation}}
\newcommand{\bea}{\begin{eqnarray}}
\newcommand{\eea}{\end{eqnarray}}
\newcommand{\barr}{\begin{array}}
\newcommand{\earr}{\end{array}}

\rightline{}
\rightline{astro-ph/0507053}
\rightline{UFIFT-HEP-05-12}
\vskip 1cm

\begin{center}
\ \\
\large{{\bf Running Non-Gaussianities in DBI Inflation}} 
\ \\
\ \\
\ \\
\normalsize{ Xingang Chen }
\ \\
\ \\
\small{\em Institute for Fundamental Theory \\ Department of Physics,
University of Florida, FL 32611 }

\end{center}

\begin{abstract}
We study the non-Gaussianity in the simplest infrared (IR) model of
the DBI inflation. We show that the non-Gaussianity in such a model
is compatible with the current observational bound, and is within the
sensitivity of
future experiments. We also discuss the scale dependence of the
non-Gaussianity. In the DBI inflation, such a feature can be
used as a probe to the properties of the background geometry of the
extra dimensions or internal space.
\end{abstract}

\setcounter{page}{0}
\thispagestyle{empty}
\maketitle

\eject

\vfill

\baselineskip=18pt

\section{Introduction}
Recently the DBI
inflation\cite{Silverstein:2003hf,Alishahiha:2004eh,Chen:2004gc,Chen:2005ad}
is proposed as an alternative to
the slow-roll inflation. The motivation comes from the non-genericness
of flat potentials. In slow-roll inflation, the required flatness
is described by the slow-roll conditions and is necessary to hold the
inflaton on the top of the potential for a sufficiently long time, and
at the same time producing a scale-invariant spectrum for density
perturbations. In
DBI inflation, one starts with a
generic steep potential, but in the mean while considers a warped
background. The inflaton can move slowly
on the steep potential because a
large warping can give rise to a stringent causality constraint on the
speed-limit of the inflaton. Despite of the slowness of the inflaton
coordinate speed, it is highly relativistic. Such inflationary models
are interesting in situations where warped space are common, but not
flat potentials.

There are two types of DBI inflation models. 
In the UV model\cite{Silverstein:2003hf,Alishahiha:2004eh}, the
inflaton slides down the potential from the UV
side of the warped space to the IR end. This results in a power law
inflation when the scale of the potential is high enough. In the IR
model\cite{Chen:2004gc,Chen:2005ad}, the inflaton is originally
trapped in the IR region through some
sort of phase transition and then rolls out from the IR to UV
side. The
resulting inflation is exponential and the potential scale is flexible.

As for many other inflationary models, when confronted with
experiments,
it is generally a combination of different properties that will pin
down
some particular models. These properties include the scalar and tensor
spectral indices,
the running of these indices and the primordial non-Gaussianity. 
In this paper,
we will be most
interested in the non-Gaussianity from the three-point correlation
functions of the scalar fluctuations. This is a function of three
momenta forming a triangle.
The property includes the overall order of
magnitude of the non-Gaussianity, its dependence on the shape of the
momentum triangle, and on the overall size of the triangle.

The slow-roll inflation usually
gives very low
non-Gaussianity\cite{Maldacena:2002vr,Gangui:1993tt,Acquaviva:2002ud},
because in the leading order the quantum
fluctuations are generated by free fields in the dS
background. However in the DBI inflation, the
causality constraint in the kinetic term introduces non-linear
interactions among different momentum modes of the scalar field. It is
therefore important to study
the level of non-Gaussianities of such models. 

The dependence on the shape of the momentum triangle is generally
complicated and model-dependent\cite{Babich:2004gb}. If such
information can be extracted
from the CMB observations, it can be very useful to distinguish
different models. The overall magnitude of the non-Gaussianity can
usually be defined when the triangle takes a specific shape, for
example, when the triangle becomes equilateral.

The dependence on the size of the triangle is relatively weaker. This
follows from the scale invariance of the inflation. However in
realistic inflation models such a scale invariance is generally
broken. So the non-Gaussianity is also titled. As we will see,
this running behavior can also be used to distinguish different
models, especially for models having similar dependence on the
triangle shape.

In Ref.~\cite{Alishahiha:2004eh} the
non-Gaussianity of a UV DBI inflation model has been
studied. There the observational bound constrains both the
non-Gaussianity and the value of the inflaton field. The required
field range and the corresponding non-Gaussianity and tensor modes
were discussed in \cite{Alishahiha:2004eh}. They can be tested in the
future experiments. In Sec.~\ref{SecIR}, we will study an IR model and
show that the non-Gaussianity here is almost independent of the
inflationary energy scale.
The predicted range of the non-Gaussianity is also within the
observational ability. 
In Sec.~\ref{SecUV}, we make comparisons
with the UV model and some slow-roll models on several interesting
aspects. In these two sections, 
we will also emphasize
the scale dependence of the non-Gaussianity, namely the dependence on
the size of the momentum triangle. This
dependence is determined by the geometry of the warped
space scanned by the inflaton during inflation, thus carries
information of the background geometry of the internal space or extra
dimensions.

In this paper, we restrict ourself to the simplest model. 
By simplest, we choose parameters so that
the effective field theory works. Complications, or sometimes
improvements, may come at least in two occasions\cite{Chen:2005ad} --
where the
red-shifted string scale is too low so that stringy effects become
significant, or where the relativistic reheating is happening in a
relatively deep 
warped space so the cosmological rescaling\cite{Chen:2004hu} takes
effect. We discuss
the first in Sec.~\ref{SecDis}.

\section{The IR model}
\label{SecIR}
In this section, we study the non-Gaussianity in the simplest IR
DBI inflation model. We begin with a brief review on the
model. Details can be found in Refs.~\cite{Chen:2005ad,Chen:2004gc}.

The inflaton potential is parameterized as
\bea
V = V_0 - \half m^2 \phi^2 = V_0 - \half \beta H^2 \phi^2 ~,
\label{PotV}
\eea
where the Hubble parameter $H$ is approximately a constant. 
In many inflationary models, there is always
naturally a contribution to the potential with $|\beta| \sim 1$. In
these models, such a
potential is too steep to support a long period of slow-roll
inflation. This is
the well-known $\eta$-problem which plagues the slow-roll
inflation\cite{Lyth:1998xn}.

However it is shown\cite{Chen:2005ad} that, with warped space, the
DBI
inflation can happen for both
small and large $\beta$ $(\beta >0)$.  
It generates a scale-invariant spectrum for
the density perturbations with a tilt independent of the parameter
$\beta$. In this case the steepness of the potential
does not play such an important role. The inflaton stays on the
potential due to a warping in the internal space,
\bea
ds^2 \propto 
\frac{\phi^2}{\lambda} ds_4^2 + \frac{\lambda}{\phi^2} d\phi^2
~,
\label{WarpedSpace}
\eea
where $ds_4^2 \equiv g_{\mu\nu} dx^\mu dx^\nu$ is the metric of the
four-dimensional space-time and
$\lambda$ is a dimensionless parameter.
The inflaton $\phi$ and the parameter $\lambda$ are related to the
notations of Refs.~\cite{Chen:2004gc,Chen:2005ad} by $\phi \equiv r
\sqrt{T_3}$ and
$\lambda \equiv
T_3 R^4 \sim N$. The $\lambda$ has the same order of magnitude as
the effective background charge $N$ of the warped space, and
characterizes the strength of the background.
The low-energy dynamics is described by the DBI-CS action
\bea
S = \frac{\mpl^2}{2} \int d^4 x \sqrt{-g} R
- \int d^4x \sqrt{-g} \left( \frac{\phi^4}{\lambda} 
\sqrt{ 1+ \frac{\lambda}{\phi^4} g^{\mu\nu} \partial_{\mu}\phi
\partial_{\nu}\phi}~ - \frac{\phi^4}{\lambda} + V(\phi) \right) ~.
\label{Action}
\eea
In the non-relativistic limit, this action reduces to the usual
minimal form.

We start the inflaton near $\phi \sim 0$ through a phase
transition.\footnote{This initial condition can be naturally obtained
without
tuning in e.g.~a scenario of Refs.~\cite{Chen:2004gc,Chen:2005ad}.}
Without the warped space, the scalar will quickly roll
down the steep potential ($\beta \gtrsim 1$) 
and make the inflation impossible. To obtain
inflation, it is natural to expect that the speed-limit should be
nearly saturated. Indeed, solving the equations of motion, we find
\bea
\phi = - \frac{\sqrt{\lambda}}{t} + \frac{9
\sqrt{\lambda}}{2\beta^2H^2} \frac{1}{t^3} + \cdots ~,
~~~~~ t \ll -H^{-1} ~,
\label{phiDynamics}
\eea
where the time $t$ is chosen to run from $-\infty$.
The inflaton travels ultra-relativistically with a Lorentz contraction
factor
\bea
\gamma = (1- \lambda \dot \phi^2/\phi^4)^{-1/2} ~.
\label{gamma}
\eea
Nonetheless, the coordinate speed of light is very small due to the
large warping near $\phi \approx 0$, and in such a way the inflaton
achieves the ``slow-rolling''. The potential stays nearly constant
during the inflation, and we have a period of exponential expansion
with the Hubble constant $H = \dot a/a =\sqrt{V_0}/\sqrt{3}\mpl$. 
There is no
lower bound on the inflationary scale, and the approximations that $H$
is constant and dominated by the potential energy during inflation
require an upper bound on $V$\cite{Chen:2005ad},
\bea
\frac{V}{\mpl^4} \ll \frac{1}{\beta \lambda N_e} ~.
\eea
Generally speaking, this bound is not significant, since to get enough
e-foldings we only need $\lambda \gtrsim 10^4$. However for some
specific models, such as the simplest one that we focus on in this
paper, $\lambda$ is determined by density perturbations and can be
much larger.

For the case $\beta \gtrsim 1$ that we are most interested in, the
behavior (\ref{phiDynamics}) is valid for $t \ll -H^{-1}$. The lower
bound on $t$ comes from back-reactions of the relativistic
inflaton\cite{Silverstein:2003hf,Chen:2004hu,Chen:2005ad} 
and the dS space\cite{Chen:2005ad} on the warped space. These effects
will smooth out the effective geometry of
certain IR region of the warped space. Therefore even if we start the
inflaton from that region, the inflationary period cannot
be further increased in terms of the order of magnitude. For the case
that we consider here,\footnote{This corresponds to the case of a
single brane rolling out of a throat in
Refs.~\cite{Chen:2004gc,Chen:2005ad}.} the strongest lower
bound is the closed
string creation from the dS background. This gives $t > 
-\sqrt{\lambda} 
H^{-1}$. So the maximum number of e-foldings in this model is of order
$\sqrt{\lambda}$. The latest e-folds $N_e$ is given by
\bea
N_e = \sqrt{\lambda} H/ \phi ~.
\label{Ne}
\eea
It has an interesting relation to the Lorentz contraction factor of
the inflaton,
\bea
\gamma \approx \beta N_e/3 ~.
\label{GammaNe}
\eea
Since the sound speed $c_s = \gamma^{-1}$, during inflation, $c_s \ll
1$.\footnote{So we
cannot use the results of Ref.~\cite{Seery:2005wm} 
where the assumption has been
made that $c_s$ departs from unity by a quantity much less than one.}

To study the perturbations, we expand the inflaton
around the above background,
\bea
\phi = -\frac{\sqrt{\lambda}}{t} + \alpha ({\bf x},t) ~.
\label{phiExpan}
\eea
It is useful to define the parameter $\zeta$,
\bea
\zeta \equiv H\frac{\alpha}{\dot \phi} + \Phi ~.
\eea
(It is shown\cite{Chen:2005ad} that the scalar metric perturbation
$\Phi$
is negligible in this model.) The parameter $\zeta$ is useful because
it remains constant after the
corresponding mode exits the horizon\cite{Mukhanov:1990me}. This can
be seen by looking at
the exact equation of motion for the linear
perturbations\cite{Garriga:1999vw}
\bea
v'' - \gamma^{-2} \nabla^2 v - \frac{z''}{z} v = 0 ~.
\label{Eomv}
\eea
The variable $v$ is defined by $v \equiv z \zeta$ with
$z \approx a \dot \phi \gamma^{3/2} H^{-1}$ in this case.
The prime in this equation denotes the derivative with respect to the
conformal time $\eta$ defined by $d\eta \equiv dt/a(t)$.
The horizon exit for a mode $\bk$
happens when $k/a \ll H \gamma$. 
The $\gamma$ factor comes in due to the
relativistic effect. It reduces the horizon size by a
factor of the sound speed $\gamma^{-1}$.
It is easy to see that in this limit
Eq.~(\ref{Eomv}) becomes $v''_\bk /v_\bk = z''/z$ and has the solution
$v_\bk = z \cdot {\rm const.}$,
so that $\zeta_\bk = {\rm const.}$. 
Therefore we will calculate the primordial scalar
correlation functions in terms of the correlation
functions of $\zeta$'s and evaluate it after the horizon crossing.
Decompose the Fourier modes of $\zeta$ as 
\bea
\zeta_\bk =
\zeta^{\rm cl}_\bk a_\bk + \zeta^{\rm cl*}_{-\bk} a^\dagger_{-\bk}
\eea
with the commutation relations 
\bea
[a_{\bf k},
a_{\bf k'}^\dagger] = (2\pi)^3 \delta^3({\bf k}-{\bf k'}) ~.
\eea
The explicit form of the classical solution $\zeta^{\rm cl}_\bk$ 
can be worked out
by examining two different
limits in Eq.~(\ref{Eomv}). When modes are well within the horizon,
$k/a \gg H \gamma$,
\bea
\zeta^{\rm cl}_\bk 
\approx -\frac{H^2}{\dot \phi} \frac{1}{\sqrt{2k^3}}
\frac{k \eta}{\gamma} e^{-i k\eta/\gamma} ~.
\label{zetaSol1}
\eea
When modes are far outside of the horizon, $k/a \ll H \gamma$, 
\bea
\zeta^{\rm cl}_\bk 
\approx i \frac{H_*^2}{\dot \phi_* \sqrt{2k^3}} ~,
\label{zetaSol2}
\eea
where the subscript ${}_*$ indicates the variable be evaluated at the
horizon crossing.

Equivalently, we can also first calculate the correlation functions of
$\alpha$ using the decomposition 
\bea
\alpha_k = u_{\bf k} a_{\bf k} + 
u_{-{\bf k}}^* a_{-{\bf k}}^\dagger ~.
\label{DecompAlpha}
\eea 
The function $u_{\bf k}(\eta)$ is the usual classical solution of the
scalar fluctuations in the dS background,
\bea
u_{\bf k}(\eta) = \frac{H}{\sqrt{2k^3}} e^{-i k\eta/\gamma} \left( i -
\frac{k\eta}{\gamma} \right) ~,
\label{uSol}
\eea
except for the presence of the $\gamma$ factors for the reason
that we
have mentioned below (\ref{Eomv}).
We then use the relation $\zeta \approx H
\alpha/\dot \phi$ to convert $\alpha$ to $\zeta$, evaluating $H$ and
$\dot \phi$ at the horizon crossing. Since the non-Gaussianity in this
model will turn out to be much larger than that in the minimal
slow-roll
model\cite{Maldacena:2002vr}, we only need consider the leading
order of $\zeta$\cite{Creminelli:2003iq,Alishahiha:2004eh}. 
With this prescription, it is
easy to see that these two methods are the same since
Eqs.~(\ref{zetaSol1}), (\ref{zetaSol2}) and (\ref{uSol}) are connected
by the same relation.

We first calculate the two-point function. This function contains
information on the density perturbations.
\bea
\langle \zeta_{\bk_1} \zeta_{\bk_2} \rangle = (2\pi)^5 
\delta^3 (\bk_1 + \bk_2) \cPR \frac{1}{2k_1^3} ~,
\eea
where $\cPR$ is the spectral density
\bea
\cPR = \frac{H_*^4}{(2\pi)^2 \dot \phi_*^2} ~.
\label{SpDen}
\eea

The calculation of the three-point function is the same
as that in the UV model\cite{Alishahiha:2004eh}. We plug the expansion
(\ref{phiExpan})
into the Lagrangian and read off the cubic terms of the scalar field
fluctuations,
\bea
{\cal L}_3 &=& a^3 \left[ \frac{\lambda \dot \phi \gamma^5}{2\phi^4}
\dot \alpha^3 - \frac{\lambda \dot \phi \gamma^3}{2\phi^4 a^2} (\nabla
\alpha)^2 \dot \alpha \right. \cr
&+& \left.
\frac{\lambda \dot \phi^2 \gamma^3}{\phi^5 a^2} (\nabla \alpha)^2
\alpha
+ \left( \frac{5\lambda \dot \phi^3 \gamma^3}{\phi^6} +
\frac{6\lambda^2 \dot \phi^5 \gamma^5}{\phi^{10}} \right) \dot \alpha
\alpha^2 
- \frac{3\lambda \dot \phi^2 \gamma^5}{\phi^5} \dot \alpha^2 \alpha
\right. \cr 
&-& \left.
\left( \frac{2 \dot \phi^2 \gamma^3}{\phi^3} + 
\frac{4\lambda^2 \dot \phi^6 \gamma^5}{\phi^{11}}
-(1-\frac{1}{\gamma}) \frac{4\phi}{\lambda} \right) \alpha^3
\right] ~.
\label{L3all}
\eea
Using
(\ref{phiDynamics}) we can estimate that the first line in
(\ref{L3all}) are the leading terms for $t \ll -H^{-1}$.\footnote{We
can also check the magnitude of the expansion parameter used for
$\cL$, which is $\gamma^2 \lambda \dot \phi \dot \alpha/\phi^4
\sim \gamma^2 \frac{H^2}{\dot \phi}$ near the horizon crossing. This
is small since the second
factor is related to the density perturbations and has to be very
small according to the COBE normalization.} 
So to the first
order\cite{Maldacena:2002vr},
\bea
\langle \zeta^3(t) \rangle =-i\int_{t_0}^t dt' \langle [ \zeta^3(t),
H_{\rm int}(t') ] \rangle ~,
\label{zeta3}
\eea
where the leading terms of the interaction Hamiltonian are
\bea
H_{\rm int} = -a^3 \frac{\lambda \dot \phi \gamma^5}{2\phi^4}
\int d^3{\bf x}
\left[ \dot \alpha^3 - \frac{(\nabla \alpha)^2 \dot
\alpha}{\gamma^2 a^2} \right] ~.
\label{Hint}
\eea
The lower limit $t_0$ in the integration (\ref{zeta3}) is some early
time when the modes are still well within the Hubble horizon. The
modes are rapidly oscillating at that time and average to zero. This
effect can also be captured by adding a damping term. The upper limit
$t$ is taken to be the time of several e-folds after the horizon
crossing. In terms of the conformal time $\eta \equiv - a^{-1}
H^{-1}$, the integration range can be taken from
$-\infty$ to a value where $\eta k/\gamma \ll 1$, or effectively to 0.

With all the prescription given above, using (\ref{DecompAlpha}),
(\ref{uSol}), (\ref{zeta3}) and (\ref{Hint}),
it is straightforward to work out the
three-point correlation function
\bea
&&\langle \zeta_{\bf k_1} \zeta_{\bf k_2} \zeta_{\bf k_3} \rangle \cr 
&=& \frac{i}{16} (2\pi)^7 ~\delta^3 (\sum_i \bk_i) 
~\cPR^2~
\frac{1}{\prod_i k_i^3} \cr
&&\times \int_{-\infty}^0 d\eta~ e^{i k_t \eta/\gamma} 
\left[ \frac{6\eta^2}{\gamma} k_1^2 k_2^2 k_3^2 - 2 \gamma k_3^2
(\bk_1\cdot\bk_2) \left( i+ \frac{k_1\eta}{\gamma} \right)
\left( i+ \frac{k_2\eta}{\gamma} \right) + {\rm perm.} \right] \cr
&&+ ~{\rm c.c.} ~,
\eea
where $k_t = k_1 + k_2 + k_3$ and Eq.~(\ref{phiDynamics}) has been
used to combine factors of $\phi$ and $\dot \phi$ into factors of 
$\cPR$. 
Since $\cPR$ remains constant after the horizon crossing, we pull
it to the front of the integration.
The ``perm.''~indicates two other terms with the same
structure as the last term but permutation of indices 1, 2 and 3.
The ``c.c.''~stands for the complex conjugate of all terms.
The damping effect for the early time can be added by
replacing $\eta \rightarrow \eta (1-i \epsilon)$. Performing the
integral we get\footnote{A more general analyses is in
progress\cite{CHS}.}
\bea
\langle \zeta_{\bf k_1} \zeta_{\bf k_2} \zeta_{\bf k_3} \rangle
= (2 \pi)^7 ~\delta^3(\sum_i \bk_i) ~\cPR^2~ F(\bk_1,\bk_2,\bk_3) ~,
\label{ThreePoint}
\eea
where the form factor is
\bea
F =
\frac{\gamma^2}{4 k_t^3 \prod_i k_i^3}
\left[ -6 k_1^2 k_2^2 k_3^2 + k_3^2 (\bk_1 \cdot \bk_2) 
\left( 2 k_1 k_2 - k_3 k_t + 2 k_t^2 \right) + {\rm perm.} \right] ~.
\label{FIR}
\eea

There are two interesting limits of the shape of the momentum
triangle. When one side of the
triangle is very small, for example, $k_1 \approx 0$ and $k_2 =
k_3$, the form factor goes as 
\bea
F \approx -\frac{11\gamma^2}{16 k_1 k_2^5} ~.
\eea
When the triangle becomes equilateral, $k_1=k_2=k_3=k$,
\bea
F = -\frac{7 \gamma^2}{ 24 k^6} ~.
\label{FeqforSL}
\eea

The non-Gaussianity of the CMB in the WMAP observations is analyzed by
assuming the simple ansatz\cite{Komatsu:2001rj,Komatsu:2003fd}
\bea
\zeta = \zeta_L - \frac{3}{5} f_{NL} \left( \zeta_L^2 - \langle
\zeta_L^2 \rangle \right) ~,
\eea
where $\zeta_L$ is the linear Gaussian part of the perturbations, and
$f_{NL}$ parameterizes the level of the non-Gaussianity. Using
$\zeta^2_L(\bk) = \int \frac{d^3 \bp}{(2\pi)^3} \zeta_L (\bp)
\zeta_L(\bk-\bp)$ and $\langle \zeta_L^2 (\bx) \rangle =
\frac{H_*^2}{\dot \phi_*^2} \int \frac{d^3 \bp}{(2\pi)^3} |u_\bp|^2$,
this assumption leads to
\bea
F =  - f_{NL}
\frac{3 \sum_i k_i^3}{10 \prod_i k_i^3} ~.
\eea
The triangle shape dependence is very different from that in the
DBI
inflation case. For example, for $k_1 \approx 0$ and $k_2 =
k_3$,
\bea
F \approx
-f_{NL} \frac{3}{5 k_1^3 k_2^3} ~.
\eea
As pointed out in Ref.~\cite{Babich:2004gb}, this feature can be used
to distinguish
different models, for example, by plotting the function $k_1^2 k_2^2
k_3^2 F$ with one side fixed ($k_3=1$).
However the momentum dependence for the equilateral case is the same,
\bea
F = - f_{NL} \frac{9}{10 k^6} ~.
\eea
We can therefore use the $f_{NL}$ in the equilateral case to compare
with the existing experimental analyses. From (\ref{FeqforSL}) we get
\bea
f_{NL} \approx 0.32 \gamma^2 ~.
\label{fNLgamma}
\eea

The momentum dependence in (\ref{FIR}) and the relation
(\ref{fNLgamma}) are the same as in the UV
model\cite{Alishahiha:2004eh}. The differences
lie in its evaluation under the zero-mode background evolution 
discussed in the beginning of this section.
Using the relation (\ref{GammaNe}), we find
\bea
f_{NL} \approx 0.036 \beta^2 N_e^2
\label{fNLNe}
\eea
for this simplest IR model.
Depending on the energy scale of the inflation, 
the CMB can correspond to an $N_e$
ranging from 30 to 60. For $1/2 \lesssim \beta \lesssim 2$, we have $8
\lesssim f_{NL} \lesssim 518$. 
For example, for $N_e \approx 50$ and $\beta
\approx 1$, we have $f_{NL} \approx 90$. This is compatible with the
current observational bound\cite{Komatsu:2003fd}
\bea
|f_{NL}| \lesssim 100 ~.
\label{fNLexp}
\eea
The WMAP will eventually reach a sensitivity $|f_{NL}| \lesssim 20$, 
and
the Planck $|f_{NL}| \lesssim 5$. So the non-Gaussianity in this model
is also within the sensitivity of
future experiments. We also note here that the dependence of
Eq.~(\ref{fNLNe}) on the inflationary energy scale is very weak (only
through $N_e$) comparing to the UV model that we will discuss in
the next section.

The triangle size dependence of the non-Gaussianity also provides
interesting information.
We are interested in the equilateral case since the momentum
dependence is the same for different models.
The size dependence in Eq.~(\ref{ThreePoint}) comes from two
different factors. The first is from $\cPR$. This
is due to the tilt of the density perturbations, so we are more
interested in the second factor coming from $F(\bk_1,\bk_2,\bk_3)$,
namely (\ref{fNLNe}). This has a logarithmic dependence on the
wavenumber.
Analogous to the spectral index, we can define an index parameterizing
the running of the overall non-Gaussianity,
\bea
n_{NG}-1 \equiv \frac{\partial \ln |f_{NL}|}{\partial \ln k} 
= - \frac{\partial \ln |f_{NL}|}{\partial N_e} 
\approx -\frac{2}{N_e} ~.
\label{nNGIR}
\eea
This tilt is directly related to the background geometry of the
internal space or extra dimensions which the inflaton scans through
during the inflation. To see this,
it is instructive to consider a different background geometry. 

Suppose
that a portion of a warped space in the IR side has a constant warp
factor,
\bea
ds^2 \propto f_0^2 ds_4^2 + f_0^{-2} d\phi^2 ~,
\label{WarpedSpace2}
\eea
where $f_0$ is a constant. Consider that the inflation is caused by
the speed-limit constraint when the inflaton moves out of this part of
the warped space. Note that the existing explicit examples of the
string theory flux compactification usually gives geometries of the
type of (\ref{WarpedSpace}) as in
Refs.~\cite{Klebanov:2000hb,Giddings:2001yu,Kachru:2003aw}, but not
(\ref{WarpedSpace2}). But we
consider this anyway just for the sake of comparison.

The inflaton dynamics can be similarly solved,
\bea
\phi = f_0^2 t + \frac{9f_0^2}{2 \beta^2 H^2 t} + \cdots ~, 
~~~~~ t \gg H^{-1} ~,
\eea
with the Lorentz contraction factor
\bea
\gamma \approx \frac{\beta H t}{3} 
\approx \frac{\beta}{3} (N_{\rm tot} - N_e) ~,
\eea
where $N_{\rm tot}$ is the total number of e-folds. Repeating the
previous procedures, we get a similar interaction Hamiltonian
\bea
H_{\rm int} = -a^3 \frac{\dot \phi \gamma^5}{2 f_0^4}
\int d^3{\bf x}
\left[ \dot \alpha^3 - \frac{(\nabla \alpha)^2 \dot
\alpha}{\gamma^2 a^2} \right] ~.
\label{Hint2}
\eea
and the same three-point function (\ref{ThreePoint}) and
(\ref{FIR}). But the relation between $\gamma$ and $N_e$ is different
from the previous case. For the equilateral case, we have
\bea
f_{NL} \approx 0.036 ~\beta^2 (N_{\rm tot} -N_e)^2 ~,
\eea
and the running index for the non-Gaussianity is 
\bea
n_{NG}-1 \approx \frac{2}{N_{\rm tot} - N_e} ~.
\eea
This has an opposite sign to Eq.~(\ref{nNGIR}).

So we see that, in the DBI inflation, the measurement of the
dependence of the
non-Gaussianity on $N_e$ encodes the relation between $\gamma$ and
$N_e$, and thus can provide information on the background
geometry of the internal space or extra dimensions. There are two
factors that can change $\gamma$. The first is the falling-down of the
potential and it always tends to increase $\gamma$ as $N_e$
decreases. The second is the shape of the background geometry. For the
case of (\ref{WarpedSpace}), the inflaton moves from the IR to the UV
side of the warped space, the causality constraint is weaker for later
time. This effect wins over the effect coming from the
potential, resulting in a negative $n_{NG}-1$. For the case of
the flat geometry (\ref{WarpedSpace2}), we only have the first factor
and hence we get a positive $n_{NG}-1$. In the next section, we will
discuss the UV model, where the two factors add up and the running
of the non-Gaussianity is much stronger.

\section{The UV model and some slow-roll models}
\label{SecUV}
In this section, we review aspects of the UV DBI
inflation model and slow-roll inflation models, and make some
comparison. 

In the UV model\cite{Silverstein:2003hf,Alishahiha:2004eh}, the
inflaton travels from the UV side of the warped
space to the IR side under the potential
\bea
V = \half m^2 \phi^2 ~.
\eea
The inflaton dynamics can be found as
\bea
\phi = \frac{\sqrt{\lambda}}{t} - \frac{\lambda^{3/2}}{4m^2 \mpl^2
t^5} + \cdots ~.
\label{phiDynUV}
\eea
The Hubble parameter $H \approx p/t$ is no longer a constant and the
inflation is in a
power law $a(t) \propto t^p$, where $p \approx
\sqrt{\frac{\lambda}{6}} \frac{m}{\mpl}$. The inflation happens
if the inflaton mass satisfies $m \gg \mpl/\sqrt{\lambda}$~,
i.e.~when $p\gg 1$.

Using (\ref{gamma}) and (\ref{phiDynUV}), we can express the Lorentz
contraction factor as
\bea
\gamma \approx 2p \frac{\mpl^2}{\phi^2}
\label{gammaphiUV}
\eea
From this equation, one can see that there is a tension between having
a small $\gamma$ and a large $p$. According to (\ref{fNLgamma}) and
(\ref{fNLexp}), in the CMB region, $\gamma \lesssim 18$. Normally
$\phi \ll \mpl$ for large $\lambda$.\footnote{For example, if we
require
$r<R$ in terms of a brane moving in a warped space with a
characteristic length scale $R$.} In this case, Eq.~(\ref{gammaphiUV})
indicates that the non-Gaussianity is too big to be compatible
with the observation. However if one considers a large
scalar vev $\phi \gtrsim \mpl$, it is still possible to satisfy the
observational bound\cite{Alishahiha:2004eh}.\footnote{For example, if
we consider a large
number, $\sqrt{\lambda}$, of branes stick together. This increases
$\phi$ by increasing the effective brane tension $T_3$ through $\phi =
r\sqrt{T_3}$, and may be possible to make $\phi \gtrsim \mpl$.} 
For example, taking $\phi \sim \mpl$, the
resulting non-Gaussianity together with the tensor modes can in
principle be detected in future experiments.

We denote the end point of the inflation as
$\phi_f$, which can come from having a warped space with a modest
total warping, or can be caused by the back-reaction of the
relativistic inflaton. The relation between $\gamma$ and $N_e$ is now
\bea
\gamma \approx \frac{2p \mpl^2}{\phi_f^2}~ e^{-2 N_e/p} ~.
\eea
So we see that the non-Gaussianity
\bea
f_{NL} \approx 1.3 \frac{p^2 \mpl^4}{\phi^4}
\approx 1.3 \frac{p^2 \mpl^4}{\phi_f^4} e^{-4N_e/p}
\eea
grows exponentially as
$N_e$ decreases. As we mentioned, this reflects
the fact that the inflaton is traveling towards a region of higher
warping.
In terms of the
index $n_{NG}$,
\bea
n_{NG}-1 \approx \frac{4}{p} ~.
\eea
Comparing to the IR case (\ref{nNGIR}), 
it has an opposite sign and
bigger magnitude (for $\phi \sim \mpl$).

\medskip
Unlike the direct scalar interactions that we have seen in the DBI
inflation,
the leading non-Gaussianity in the
simplest model of the slow-roll inflation originates from the
non-linearities of the Einstein action and the flat
potential.
Calculations\cite{Maldacena:2002vr,Gangui:1993tt,Acquaviva:2002ud}
show that $f_{NL}$ is in the same order of the
slow-roll parameters, and therefore is unobservably small. One can
consider non-minimal cases of the slow-roll models, for
example, by
adding a correction term $\frac{1}{8M^4} (\partial_\mu \phi
\partial^\mu \phi)^2$\cite{Shiu:2002kg}, where $M$ is the energy scale
of new physics.
Interestingly, this term gives a
non-Gaussianity\cite{Creminelli:2003iq} with the same shape
dependence as in the DBI inflation models that we just discussed. The
difference is the overall magnitude -- the $\gamma^2$ in
(\ref{FIR}) is replaced by $\dot \phi^2/M^4$. The running of $f_{NL}$
depends on
the shape of the potential. For example, using
the slow-roll relations
$\dot \phi \approx -V'(\phi)/3H$ and $\phi=\phi_m e^{\mp\beta N_e/3}$
($\phi_m$ is the end point of the inflation), 
we have $n_{NG}-1 = \pm 2\beta/3$ for
the potential $V=V_0 \mp \half \beta H^2 \phi^2$.
But in any case,
since $\dot \phi^2/M^4$ has
to be less than one to justify the neglect of higher order
corrections,
$f_{NL}$ has to be less than one and is also too small to be observed.

\section{Discussions}
\label{SecDis}
We have studied the non-Gaussianity in the simplest IR model of the
DBI inflation and shown that it satisfies the current
experimental bound and provides interesting predictions for future
test. In this section, we would like to discuss a non-minimal
case\cite{Chen:2005ad}.

Such a case deserves further investigations because there are at least
two aspects of the simplest model that may not be fully
satisfying. The first comes form the fitting of the density
perturbations. Using (\ref{SpDen}) and (\ref{Ne}), we find the density
perturbations on the CMB scale
\bea
\delta_H \approx \frac{N_e^2}{5\pi \sqrt{\lambda}} ~.
\label{deltaH1}
\eea
The COBE normalization $\delta_H \approx 1.9\times 10^{-5}$ at $N_e
\sim 60$ requires $\lambda \sim 10^{14}$.

Whether such a number is natural depends on the fundamental physics
which realizes the model. The brane inflation\cite{Dvali:1998pa} in
warped string
compactification\cite{Klebanov:2000hb,Giddings:2001yu,Kachru:2003aw}
is a natural place to realize the DBI
inflation\cite{Chen:2004gc,Chen:2005ad}, where the brane position
moving in the warped extra
dimensions plays the role of the scalar
inflaton moving in the warped internal space, as in the slow-roll
models\cite{Dvali:2001fw,Burgess:2001fx,Kachru:2003sx,Shandera:2003gx}.
The inflationary energy
can be provided by anti-branes or simply by a moduli potential. In
this
type of scenario, $\lambda$ is in the order of the effective 
background charge
of the warped throat $N$. In this context, this required number
is extremely
large\footnote{Although, a somewhat different point of view is to
regard it as a requirement that the characteristic length scale of the
throat
be $\cO(10^3 \sim 10^4)$ in string units, and so, in general, the
tuning of $\lambda$ may be different from the tuning of $N$. I thank
Eva Silverstein for pointing it out to me.

Other discussions
can be found in
Refs.~\cite{Alishahiha:2004eh,Chen:2004gc,Chen:2005ad}.
For example, the effect of the cosmological
rescaling can greatly reduce $\delta_H$. This happens if the reheating
happens in a warped space with a relatively large
warping\cite{Chen:2005ad,Chen:2004hu}.}
\cite{Alishahiha:2004eh,Chen:2004gc,Chen:2005ad,Firouzjahi:2005dh}.

The second concern is that the tilt of the density perturbations,
\bea
n_s -1 \approx -\frac{4}{N_e}
\label{ns1}
\eea
may be too large to fit the observations.

Interestingly these two concerns may be addressed at the same time
without adding any new features to the model. We need look
more carefully at the validity region of our field theory
analyses of the quantum fluctuations in dS space. There are three
regions of $N_e$ that are interesting given
a $\lambda$ (which now is not necessarily large).
For a large $N_e$, the brane (or inflaton) is in the deep infrared
region of the warped space. The energy density of the scalar quantum
fluctuations has to be less than the red-shifted string scale in order
for the field theory analyses to hold. The former can be estimated
as (for an instantaneous observer moving with the brane)
\bea
\gamma^2 \Delta \phi^2/\Delta x^2 ~,
\label{ScalarEDen}
\eea
where the horizon size $\Delta x \sim \gamma^{-1} H^{-1}$ and the
Quantum
fluctuation $\Delta \phi \sim H$. The factor of $\gamma$
comes from the restoration of the Lorentz contraction factor for the
moving observer on the brane. The latter is $T_3 h^4 =
\phi^4/\lambda$, where the $h$ is the warp factor at $\phi$. Using
(\ref{Ne}) and (\ref{GammaNe}), we get the critical
e-folding
\bea
N_c \sim \frac{\lambda^{1/8}}{\beta^{1/2}} ~.
\eea
The first region is $N_e < N_c$, where the field theory
holds. This is the region we have been studying in
this paper. In the second region, $N_e > N_c$, the stringy
fluctuations become significant. We can no longer use the DBI action,
which is a low energy
field theory approximation, to calculate the quantum fluctuations. But
the inflation still
proceeds as long as the warped background is strong enough, since the
only thing important for the zero-mode inflaton dynamics is the
speed-limit causality constraint. We need
a different way to estimate the quantum fluctuations. 
A rough estimate goes as follows.
We assume the
part of energy that goes into the scalar excitations to
be $\cO(T_3 h^4)$, and the rest is used to excite strings.
From (\ref{ScalarEDen}), this gives $\Delta \phi \sim
\sqrt{\lambda} H/N_e^2 \gamma^2$.
While the strings get diluted in the later spatial expansion, the
position dependent time delay $\delta t$ caused by the
scalar fluctuations remains constant. Therefore, at the reheating, we
can still
approximate the density perturbations using the time delay without
considering the diluted stringy excitations. So we have
\bea
\delta_H = \frac{2}{5} H \delta t \approx \frac{2}{5} H \Delta
\phi/\dot \phi \sim \frac{2}{5\gamma^2} \sim
\frac{18}{5\beta^2 N_e^2} ~.
\label{deltaH2}
\eea
Note that this is very different from (\ref{deltaH1}). It is
independent of $\lambda$ (therefore no need to have a large
$\lambda$) and has an opposite running 
\bea
n_s-1 \sim \frac{4}{N_e} ~.
\label{ns2}
\eea
The third region of $N_e$ is therefore the transition region between
(\ref{ns1}) and (\ref{ns2}). This region has two interesting
properties -- it should have smaller $|n_s-1|$ but larger $|d n_s/d\ln
k|$ because it has to connect (\ref{ns1}) and (\ref{ns2}) in a short
range. The non-Gaussianity feature for $N_e \gtrsim N_c$, including
its dependence on the background geometry, should be
very interesting and remains a challenge for future studies.

\acknowledgments
I would like to thank Eva Silverstein for many valuable suggestions
and discussions on the paper, and Min-xin Huang for a numerical
correction. I would also like to thank Shamit
Kachru, Renata Kallosh, Feng-Li Lin, Andrei Linde, Gary Shiu and
Richard Woodard
for helpful discussions, and the ITP at Stanford University for the
hospitality. This work was supported in part by the US Department of
Energy.


\begin{thebibliography}{}


\bibitem{Silverstein:2003hf}
  E.~Silverstein and D.~Tong,
  ``Scalar speed limits and cosmology: Acceleration from D-cceleration,''
  Phys.\ Rev.\ D {\bf 70}, 103505 (2004)
  [arXiv:hep-th/0310221].


\bibitem{Alishahiha:2004eh}
  M.~Alishahiha, E.~Silverstein and D.~Tong,
  ``DBI in the sky,''
  Phys.\ Rev.\ D {\bf 70}, 123505 (2004)
  [arXiv:hep-th/0404084].


\bibitem{Chen:2004gc}
  X.~g.~Chen,
  ``Multi-throat brane inflation,''
  Phys.\ Rev.\ D {\bf 71}, 063506 (2005)
  [arXiv:hep-th/0408084].


\bibitem{Chen:2005ad}
  X.~g.~Chen,
  ``Inflation from warped space,''
  arXiv:hep-th/0501184.


\bibitem{Maldacena:2002vr}
  J.~Maldacena,
  ``Non-Gaussian features of primordial fluctuations in single field
  inflationary models,''
  JHEP {\bf 0305}, 013 (2003)
  [arXiv:astro-ph/0210603].


\bibitem{Gangui:1993tt}
  A.~Gangui, F.~Lucchin, S.~Matarrese and S.~Mollerach,
  Astrophys.\ J.\  {\bf 430}, 447 (1994)
  [arXiv:astro-ph/9312033].


\bibitem{Acquaviva:2002ud}
  V.~Acquaviva, N.~Bartolo, S.~Matarrese and A.~Riotto,
  ``Second-order cosmological perturbations from inflation,''
  Nucl.\ Phys.\ B {\bf 667}, 119 (2003)
  [arXiv:astro-ph/0209156].



\bibitem{Babich:2004gb}
  D.~Babich, P.~Creminelli and M.~Zaldarriaga,
  ``The shape of non-Gaussianities,''
  JCAP {\bf 0408}, 009 (2004)
  [arXiv:astro-ph/0405356].


\bibitem{Chen:2004hu}
  X.~g.~Chen,
  ``Cosmological rescaling through warped space,''
  Phys.\ Rev.\ D {\bf 71}, 026008 (2005)
  [arXiv:hep-th/0406198].


\bibitem{Lyth:1998xn}
  D.~H.~Lyth and A.~Riotto,
  ``Particle physics models of inflation and the cosmological density
  perturbation,''
  Phys.\ Rept.\  {\bf 314}, 1 (1999)
  [arXiv:hep-ph/9807278].


\bibitem{Seery:2005wm}
  D.~Seery and J.~E.~Lidsey,
  ``Primordial non-gaussianities in single field inflation,''
  JCAP {\bf 0506}, 003 (2005)
  [arXiv:astro-ph/0503692].


\bibitem{Mukhanov:1990me}
  V.~F.~Mukhanov, H.~A.~Feldman and R.~H.~Brandenberger,
  ``Theory Of Cosmological Perturbations. Part 1. Classical Perturbations. Part
  2. Quantum Theory Of Perturbations. Part 3. Extensions,''
  Phys.\ Rept.\  {\bf 215}, 203 (1992).


\bibitem{Garriga:1999vw}
  J.~Garriga and V.~F.~Mukhanov,
  ``Perturbations in k-inflation,''
  Phys.\ Lett.\ B {\bf 458}, 219 (1999)
  [arXiv:hep-th/9904176].


\bibitem{Shiu:2002kg}
  G.~Shiu and I.~Wasserman,
  ``On the signature of short distance scale in the cosmic microwave
  Phys.\ Lett.\ B {\bf 536}, 1 (2002)
  [arXiv:hep-th/0203113].


\bibitem{Creminelli:2003iq}
  P.~Creminelli,
  ``On non-gaussianities in single-field inflation,''
  JCAP {\bf 0310}, 003 (2003)
  [arXiv:astro-ph/0306122].


\bibitem{Komatsu:2001rj}
  E.~Komatsu and D.~N.~Spergel,
  ``Acoustic signatures in the primary microwave background bispectrum,''
  Phys.\ Rev.\ D {\bf 63}, 063002 (2001)
  [arXiv:astro-ph/0005036].


\bibitem{Komatsu:2003fd}
  E.~Komatsu {\it et al.},
  ``First Year Wilkinson Microwave Anisotropy Probe (WMAP) Observations: Tests
  of Gaussianity,''
  Astrophys.\ J.\ Suppl.\  {\bf 148}, 119 (2003)
  [arXiv:astro-ph/0302223].


\bibitem{Klebanov:2000hb}
I.~R.~Klebanov and M.~J.~Strassler,
``Supergravity and a confining gauge theory: Duality cascades and
chiSB-resolution of naked singularities,''
JHEP {\bf 0008}, 052 (2000)
[arXiv:hep-th/0007191].


\bibitem{Giddings:2001yu}
S.~B.~Giddings, S.~Kachru and J.~Polchinski,
``Hierarchies from fluxes in string compactifications,''
Phys.\ Rev.\ D {\bf 66}, 106006 (2002)
[arXiv:hep-th/0105097].


\bibitem{Kachru:2003aw}
S.~Kachru, R.~Kallosh, A.~Linde and S.~P.~Trivedi,
``De Sitter vacua in string theory,''
Phys.\ Rev.\ D {\bf 68}, 046005 (2003)
[arXiv:hep-th/0301240].


\bibitem{Dvali:1998pa}
G.~R.~Dvali and S.~H.~H.~Tye,
``Brane inflation,''
Phys.\ Lett.\ B {\bf 450}, 72 (1999)
[arXiv:hep-ph/9812483].


\bibitem{Dvali:2001fw}
G.~R.~Dvali, Q.~Shafi and S.~Solganik,
``D-brane inflation,''
arXiv:hep-th/0105203.


\bibitem{Burgess:2001fx}
C.~P.~Burgess, M.~Majumdar, D.~Nolte, F.~Quevedo, G.~Rajesh and R.~J.~Zhang,
``The inflationary brane-antibrane universe,''
JHEP {\bf 0107}, 047 (2001)
[arXiv:hep-th/0105204].


\bibitem{Kachru:2003sx}
S.~Kachru, R.~Kallosh, A.~Linde, J.~Maldacena, L.~McAllister and S.~P.~Trivedi,
``Towards inflation in string theory,''
JCAP {\bf 0310}, 013 (2003)
[arXiv:hep-th/0308055].


\bibitem{Shandera:2003gx}
  S.~Shandera, B.~Shlaer, H.~Stoica and S.~H.~H.~Tye,
  ``Inter-brane interactions in compact spaces and brane inflation,''
  JCAP {\bf 0402}, 013 (2004)
  [arXiv:hep-th/0311207].


\bibitem{Firouzjahi:2005dh}
  H.~Firouzjahi and S.~H.~Tye,
  ``Brane inflation and cosmic string tension in superstring theory,''
  JCAP {\bf 0503}, 009 (2005)
  [arXiv:hep-th/0501099].


\bibitem{Kachru:2002gs}
S.~Kachru, J.~Pearson and H.~Verlinde,
``Brane/flux annihilation and the string dual of a non-supersymmetric  field
theory,''
JHEP {\bf 0206}, 021 (2002)
[arXiv:hep-th/0112197].


\bibitem{DeWolfe:2004qx}
O.~DeWolfe, S.~Kachru and H.~Verlinde,
``The giant inflaton,''
JHEP {\bf 0405}, 017 (2004)
[arXiv:hep-th/0403123].

\bibitem{CHS}
X.~g.~Chen, M.~x.~Huang and G.~Shiu, in progress. 


\end{thebibliography}
\end{document}